\documentclass{elsart}
\usepackage{amsfonts}
\input tcilatex
\QQQ{Language}{
American English
}

\begin{document}

\begin{frontmatter}

\title{The Nambu-Jona Lasinio mechanism and electroweak symmetry 
breaking in the Standard Model}

\author{Edoardo Di Napoli}

\address{University of Rome "La Sapienza" P.le Aldo Moro, 5 Rome Italy}

\begin{abstract}
In this paper I examine the breaking of internal symmetries from another 
point of view showing that is possible to reproduce the electroweak  
panorama of the traditional Standard Model in a exhaustive and self consistent way. The 
result is reached applying the main futures of the old Nambu-Jona Lasinio (NJL) mechanism
to an electroweak invariant Lagrangian. In this context the use of functional formalism
for composite operators naturally leads to a different dynamical approach. While 
the Higgs mechanism acts on the Lagrangian form, a NJL like model looks directly 
at the physics of the system showing the real dynamical content hidden in the Green 
functions of the theory.        

\end{abstract}

\end{frontmatter}

\section{Introduction}

The aim of this model is to develop and analyze an alternative version of
the traditional electroweak sector of the Standard Model (SM). In the latter
the mechanism of mass ''generation'', for the gauge and matter field, is the
direct consequence of the insertion in the Lagrangian of a gauge invariant
scalar field with Yukawa coupling terms. Modifying suitably the potential
parameters, the appearance of a vacuum expectation value introduces mass
terms in the Lagrangian leading to a spontaneous symmetry breaking effect. A
different approach reaches the same results through a process that doesn't
act on the Lagrangian form but directly on the Green functions. The starting
point of the whole mechanism draws inspiration from the basic work of Nambu
and Jona Lasinio (NJL) that gave origin to the homonymous model \cite{b0}.
NJL, aware of the tight analogy with the BCS theory of superconductivity,
applied Bogoliubov's quasi particle method to the relativistic field theory.
Through a simple four fermion chirally invariant interaction, they obtained
a gap equation for an order parameter that is merely the dynamical fermion
mass. Moreover the model points out the presence of induced relativistic
bound states as much as an unavoidable logarithmic dependence on the cut-off
of the analytical results which, paradoxically, enriches the theory. In the
first part of this work the fundamental ideas of this model are presented
and then used with the combined support of the composite operators
functional formalism for effective action and 1/N expansion techniques. I
start by introducing a simplified four fermion (4-F) interaction term in a
Lagrangian invariant under the $SU_L(2)$ $\otimes $ $U_Y(1)$ symmetry group.
Using stationary condition on the effective action I find a Schwinger-Dyson
(SD) equation for the top quark propagator in the form of a self consistent
condition for its dynamical mass. After having fine tuned this condition,
setting the system in the asymmetrical phase (breaking chiral invariance), I
proceed to extract the correct Bethe-Salpeter equations. The latter show the
presence of a series of bound states playing the role of Goldstone and Higgs
bosons of the traditional SM. Going on, calculating the gauged SD equations,
it is possible to ascertain the absorption of the pseudo-Goldstone modes by
the electroweak gauge propagators. In this way mass terms are induced for W$%
^{\pm }$ and Z$^0$, while the photonic propagator remains mass free. Section
2 is a kind of toolbox in which a short explanation of the mathematical
instruments, used in the rest of the paper, is given. In section 3 I
introduce a 4-F interaction and analyze the SD equations for the fermion
propagators. Section 4 is devolved to the relativistic bound states and to
their mass spectrum. In section 5 I take in consideration the SD equation
for the gauge electroweak propagators showing the mass inducing effect
caused by the pseudo-Goldstone amplitudes. Considerations and conclusions
follow in the last section.

\section{Functional formalism toolbox}

The first problem to be resolved is the possibility of using the same
formalism to analyze order parameters (as the mass operator functions) and
bound states amplitudes. Second, but not less important, is the quest for an
equivalent approximation order for all the results obtained in the course of
the work. The functional effective action for composite operators seemed to
have all the properties we were looking for. First of all its stationarity
under variation with respect to the second Legendre variable (on physical
states) gives directly a SD equation for a local propagator that includes
all the corrections coming from the interaction terms in the Lagrangian.
Then, the second derivative respect this propagator represents the inverse
bilocal connected function of the theory, that in simple words is the
connected part of a four point particle scattering amplitude. The second
requirement on bound states is then accomplished. It remains the
approximation problem. It was evident that I should consider some kind of
cutting process so to make possible the practical handling of expressions in
terms of the most relevant contributions. Moreover the order of
approximation had to be the same for every explicit formula. In this case
the use of functional formalism points out, as natural solution, the
approximation in series of 1/N$_c$ (where N$_c$ is number of colors) showing
the high level of self consistency of this approach.

For simplicity let me introduce the functional formalism for composite
operators in case of a scalar field $\phi (x)$. As we already know, the
application of the Legendre transform to the usual generating functional $%
\mathcal{Z}(J)$, that depends on the local source $J(x)$, gives the
generator of the 1-particle irreducible graphs of the theory. If we formally
define a generating functional that also depends on a bilocal source $K(x,y)$%
\begin{eqnarray}
\mathcal{Z}(J,K) &=&\exp \left[ \frac i\hbar W(J,K)\right] =\int d\phi \
\exp \left\{ \frac i\hbar \left[ \int d^4x\ \left( \mathcal{L}(\phi )+\phi
(x)\ J(x)\right) +\right. \right.   \nonumber \\
&&\ \ \ \ \left. \left. +\iint d^4xd^4y\ \phi (x)\ K(x,y)\ \phi (y)\right]
\right\} .  \label{k1}
\end{eqnarray}
we can imagine to recover an analogous generator with a similar transform. I
will call the latter second Legendre transform, 
\begin{eqnarray}
\Gamma (\phi _c,G) &=&W(J,K)-\int d^4x\ \phi _c(x)\ J(x)-\frac 12\iint
d^4xd^4y\ \phi _c(x)\ K(x,y)\times   \nonumber  \label{k2} \\
&&\times \phi _c(y)-\frac 12\hbar \iint d^4xd^4y\ G(x,y)\ K(x,y)  \label{k2}
\end{eqnarray}
with 
\begin{eqnarray}
\frac{\delta W(J,K)}{\delta J(x)} &=&\phi _c(x)=\left\langle 0\left| \phi
\left( x\right) \right| 0\right\rangle   \nonumber \\
\frac{\delta W(J,K)}{\delta K(x,y)} &=&\frac 12\left[ \phi _c(x)\phi
_c(y)+\hbar G(x,y)\right] =\left\langle 0\right| \phi (x)\phi (y)\left|
0\right\rangle .
\end{eqnarray}
The functional so obtained is consequently the generator of the 2-particle
irreducible Green functions$\footnote{%
We call 2 point irreducible a fuction whose graph remains connected cutting
any couple of internal lines.}$ with all the internal lines expressed in
terms of the complete propagator $G\left( x,y\right) $. The word complete
here is used because $G(x,y)$ automatically takes into account the
corrections coming from the interaction terms in the Lagrangian. This
functional together with its two stationary conditions 
\begin{eqnarray}
\frac{\delta \Gamma (\phi _c,G)}{\delta \phi _c(x)} &=&-J(x)-\int d^4y\
K(x,y)\ \phi _c(y)  \label{k2b} \\
\frac{\delta \Gamma (\phi _c,G)}{\delta G(x,y)} &=&-\frac 12\hbar K(x,y). 
\nonumber
\end{eqnarray}
give the Jackiw, Cornwall and Tomboulis\cite{b1} variational method for
composite operators. On physical states, when all the sources are equal to
zero, (\ref{k2b}) become a system of equations from which is formally
possible to extract the expressions for $\phi _c$ and $G$. In particular the
latter equation, depending on self-energy, represents essentially the SD
equation for $G$. In order to extract an explicit expression for $G$ we have
to obtain a practical method that gives formulas to be handled with
approximation procedures. One of the most known is the loop expansion\cite
{b1}\cite{b3}. Referring to the usual scalar field action $S\left( \phi
\right) =\int d^4x\ \mathcal{L}\left( \phi \left( x\right) \right) $ and
defining 
\[
i\frak{D}^{-1}(\phi ;x,y)=\frac{\delta ^2S(\phi )}{\delta \phi (x)\delta
\phi (y)}=iD^{-1}(x-y)+\frac{\delta ^2S_{int}(\phi )}{\delta \phi (x)\delta
\phi (y)}
\]
with $D^{-1}$ the inverse free propagator, it is possible to write the
effective action as 
\begin{eqnarray}
\Gamma (\phi _c,G)=S(\phi _c)+\frac 12i\hbar \text{Tr}\left[ \mathrm{\ln }%
\left( DG^{-1}\right) +\frak{D}^{-1}(\phi _c)G-1\right] +\Gamma _2(\phi
_c,)G.  \label{k3}
\end{eqnarray}
An infinite number of terms are stored in $\Gamma _2$ and its definition is
possible through a slightly different version of the classic action where
the field $\phi $ is translated by the quantity $\phi _c$. This defines a
new interaction $S_{int}\left( \phi ;\phi _c\right) $ whose vertices
explicitly depend on $\phi _c$. With this premise $\Gamma _2$ is made by the
collection of all 2-particle irreducible vacuum graphs determined by a
theory with the interaction $S_{int}\left( \phi ;\phi _c\right) $ and all
the internal lines equal to $G$. These graphs are easily classifiable in a
loop series giving a coherent expression for (\ref{k3}). This argument is
equally reproducible for gauge boson and fermion fields with the exceptions
that for the latter $\psi _c=\left\langle 0\left| \psi \right|
0\right\rangle $ is identically zero (with the consequence that $\frak{D}%
^{-1}$ is equal to $D^{-1}$) and all the -$\frac 12$ factors must be
replaced with $1$. Let us imagine, now, to set the first stationary
condition such that $\phi _c=0$, the functional $W(0,K)=\frac 1i\ln Z(0,K)$
becomes the generator of the bilocal connected Green functions 
\begin{eqnarray}
W(0,K)=\stackunder{n=0}{\sum^\infty }\frac{(i)^{n-1}}{n!}G_c^{(n)}(x_1y_1;%
\ldots ;x_ny_n)K(x_1;y_1)\ldots K(x_n;y_n)  \label{k4}
\end{eqnarray}
with 
\[
G_c^{(n)}(x_1y_1;\ldots ;x_ny_n)=\frac{\delta ^nW(K)}{\delta
K(x_1;y_1)\ldots \delta K(x_n;y_n)}
\]
while $\Gamma (0,G)$ is the sum of all 2-particle irreducible (2PI) vacuum
graphs of the same theory. If we substitute now the scalar field with a
generic spinorial one, the last formula do not involve any ansatz on $\psi _c
$, but are a natural consequences of the antisymmetric character of the
field. In this scenario the combined effect of the two formula 
\begin{eqnarray*}
\frac{\delta W(K)}{\delta K(x;y)}=-G(x;y)\qquad \frac{\delta \Gamma (G)}{%
\delta G(x;y)}=K(x;y)
\end{eqnarray*}
leads to the integro differential equation 
\[
\iint d^4x\ d^4y\ \frac{\delta ^2\Gamma }{\delta G_{\alpha \beta
}(x_1;y_1)\delta G_{\gamma \delta }(x;y)}\frac{\delta ^2W}{\delta K_{\delta
\gamma }(x;y)\delta K_{\eta \theta }(x_2;y_2)}=
\]
\[
=-\left[ \delta _{\alpha \eta }\delta _{\beta \theta }\delta (x_1-x_2)\delta
(y_1-y_2)\right] 
\]
where the Greek indices are spinorial. This demonstrates that $\Gamma
^{(2)}(x_1y_1;x_2y_2)$ is the inverse of the four point connected Green
function $G_c^{(2)}(x_1y_1;x_2y_2)$. Using now the loop expansion to
calculate $\Gamma ^{(2)}$, we simply recover a compact expression for $%
G_c^{(2)}$ in momentum space 
\begin{eqnarray}
G_{c;\alpha \beta ,\gamma \delta }^{(2)}\left( p;q;P\right)  &=&G_{0;\alpha
\beta ,\gamma \delta }^{(2)}\left( p;q;P\right) +\frac 1{(2\pi )^8}\iint
d^4q^{\prime }\ d^4p^{\prime }\ G_{0;\alpha \beta ,\rho \tau }^{(2)}\left(
p;q^{\prime };P\right) \times   \nonumber \\
&&\ \ \ \times \mathsf{K}_{\rho \tau ,\sigma \eta }\left( p^{\prime
};q^{\prime };P\right) G_{c;\sigma \eta ,\gamma \delta }^{(2)}\left(
q^{\prime };q;P\right)   \label{k5}
\end{eqnarray}
where 
\begin{eqnarray*}
-i\frac{\delta ^2\Gamma _2}{\delta G\delta G}\left( p;q;P\right)  &=&\mathsf{%
K}\left( p;q;P\right)  \\
G_{0;\alpha \beta ,\gamma \delta }^{(2)}\left( p;q;P\right)  &=&(2\pi
)^4\delta (p-q)G_{\alpha \delta }(p+\eta P)G_{\gamma \beta }(p-(1-\eta )P).
\end{eqnarray*}
This is exactly the Bethe-Salpeter (BS) equation cited at the beginning of
the section. For cases in which its expression is relatively simple, this
equation is the starting point from which to detect the presence of bound
states. In the present one the expansion in series of $\frac 1{N_c}$ of the
kernel $\mathsf{K}\left( p;q;P\right) $, permits to solve it in an elegant
and self consistent way.

\section{The breaking of chiral invariance}

After the technical premises of the previous section we can enter the core
of the model. Let us consider a local $SU_L\left( 2\right) \otimes U_Y\left(
1\right) $ gauge invariant Lagrangian divided in the following groups of
terms: a pure gauge electroweak sector $\mathcal{L}_B$, a pure fermionic
sector $\mathcal{L}_F$ including all three families of quarks and leptons
and a 4-F interaction $\mathcal{L}_{4-F}$ (note that we are not including
any Higgs or Yukawa term). A sufficiently general 4-F interaction expression
could be 
\begin{eqnarray}
\mathcal{L}_{4-F}=K_{\alpha ,\beta }^{a,b}\left[ \left( \overline{\mathbf{%
\Psi }}_L^\alpha \psi _R^a\right) \left( \overline{\psi }_R^b\mathbf{\Psi }%
_L^\beta \right) \right] 
\end{eqnarray}
$\alpha $,$\beta $ and $a,b$ being respectively family and isospin indexes.
In order to be gauge invariant, the possible indexes of this expression are
constrained by the values of the weak hypercharge quantum numbers. This
implies that both the pairs $\alpha ,\beta $ and $a,b$ must be at the same
time quark like or lepton like. Moreover $a$ and $b$ must always refer to
the same isospin orientation. We need now to make a reasonable hypothesis,
on a phenomenological basis, so as to restrict the number of possible
interaction terms. A strong indication comes from the abnormal heaviness of
the quark top. It could be a significative signal of its major role in
comparison with the other lighter fermions suggesting to ignore all other
interaction terms apart from those involving it directly. This doesn't mean
that all the other terms are canceled, but that, for the moment, we consider
a simplified model with 
\begin{eqnarray}
\mathcal{L}_{4-F} &=&K_{3q,3q}^{t,t}\left[ \left( \overline{\mathbf{\Psi }}%
_L^{3q}t_R\right) \left( \overline{t}_R\mathbf{\Psi }_L^{3q}\right) \right] 
\label{k6} \\
\  &=&\frac{\mathcal{G}}{N_c}\left\{ \left( \frac 12\overline{t}t\right)
^2+\left( \frac 12\overline{t}i\gamma _5t\right) ^2+\left[ \overline{b}\frac{%
\left( 1+\gamma _5\right) }2t\ \overline{t}\frac{\left( 1-\gamma _5\right) }%
2b\right] \right\}   \nonumber
\end{eqnarray}
where $b$ indicates quark bottom and $t$ quark top. Expression (\ref{k6}) is
crucial for the determination of $\Gamma _2$ in the loop series expansion of
the effective action 
\begin{eqnarray}
\Gamma \left( G_t,G_b\right)  &=&iN_c\sum_{a=b,t}\iint d^4x\ d^4y\ \text{Tr}%
\left[ \ln S_a^{-1}(x-y)G_a(y;x)\right. +  \nonumber \\
&&\ \ \ \ \left. -S_a^{-1}(x-y)G_a(y;x)+1\right] +\Gamma _2\left(
G_t,G_b\right) 
\end{eqnarray}
where every N$_c$ comes from the sum over color for each quark loop
considered and $S_a^{-1}$ is an inverse free propagator of the theory. The
loop expansion of $\Gamma _2$ gives schematically 
\begin{eqnarray}
\Gamma _2\left( G_t,G_b\right)  &=&N_cF_{-1}\left( \text{Tr}G_t\right)
+F_0\left( \text{Tr}G_t+\text{Tr}\left( G_t;G_b\right) +\ldots \right) + 
\nonumber \\
&&\ +\frac 1{N_c}F_1\left( \text{Tr}G_t+\text{Tr}\left( G_t;G_b\right)
+\ldots \right) +\ldots   \nonumber
\end{eqnarray}
with 
\begin{eqnarray*}
F_{-1} &=&-\mathcal{G}\dint d^4x\left\{ \left[ \text{Tr}\left( \frac
12G_t(x;x)\right) \right] ^2+\left[ \text{Tr}\left( \frac 12i\gamma
_5G_t(x;x)\right) \right] ^2\right\}  \\
F_0 &=&\mathcal{G}\dint d^4x\ \left\{ \text{Tr}\left[ \left( \frac 12i\gamma
_5G_t(x;x)\right) ^2\right] +\text{Tr}\left[ \left( \frac 12G_t(x;x)\right)
^2\right] \right. + \\
&&\ +\left. \text{Tr}\left[ \frac{(1-\gamma _5)}2G_t(x;x)\frac{\left(
1+\gamma _5\right) }2G_b(x;x)\right] \right\} +\ldots 
\end{eqnarray*}
Taking in consideration the highest order in the $\frac 1{N_c}$ expansion we
finally arrive to 
\begin{eqnarray}
\Gamma \left( G_t,G_b\right)  &=&iN_C\sum_{a=b,t}\iint d^4x\ d^4y\ \text{Tr}%
\left[ \ln S_a^{-1}(x-y)G_a(y;x)\right. +  \nonumber \\
&&\ -\left. S_a^{-1}(x-y)G_a(y;x)+1\right] -\mathcal{G}\dint d^4x\
N_C\left\{ \left[ \text{Tr}\left( \frac 12G_t(x;x)\right) \right] ^2\right. +
\nonumber  \label{k7} \\
&&\ +\left. \left[ \text{Tr}\left( \frac 12i\gamma _5G_t(x;x)\right) \right]
^2\right\} .  \label{k7}
\end{eqnarray}
As we can see this expression contains terms of the same order and so its
solution will be completely self-consistent. But before starting to search
for it, is necessary to make some remarks. Let's go back to the stationary
conditions (\ref{k2b}) and analyze their meanings on the mass shell 
\begin{eqnarray}
\frac{\delta \Gamma (\phi _c,G)}{\delta \phi _c(x)} &=&0  \label{k2a} \\
\frac{\delta \Gamma (\phi _c,G)}{\delta G(x,y)} &=&0.  \nonumber
\end{eqnarray}
In the traditional SM the Lagrangian depends on an explicit complex scalar
doublet, invariant under global $SU_L(2)$ transform. The effective action is
then the first Legendre transform of the functional action $Z(J)$, meanwhile
the first stationary condition (\ref{k2a}) (the only one in this case)
determines univocally $\phi _c$. It is just a positive value of $\phi _c$,
on the mass shell, that breaks the global $SU_L(2)$ leading the electroweak
breaking. In the present case the attention is, instead, focused on the
second relation of (\ref{k2a}). The main target is to find a solution for $%
G_t$ that breaks chiral invariance spontaneously giving a dynamical value to
the mass operator. As the global symmetry breaking in the Higgs mechanism,
the chiral symmetry breaking is just a means trough which it is possible to
reach the electroweak breaking. We can now impose on (\ref{k7}) the second
of the (\ref{k2a}), calculated for the quark top propagator, remembering
that the breaking of a loop, in the derivation process, carries a
multiplicative $\frac 1{N_c}$ term 
\begin{eqnarray}
\frac{\delta \Gamma (G_t,G_b)}{\delta G_t(t;z)}=iG_t^{-1}(p)-iS_t^{-1}(p)-%
\frac 12\frac{\mathcal{G}}{(2\pi )^4}\int d^4k\ \text{Tr}\left[
G_t(k)\right] =0.  \label{k9}
\end{eqnarray}
A possible solution for $G_t$ is obtained with the help of the additional
hypothesis of linearity \`a la Hartree-Fock. Writing $%
G_t^{-1}(p)=S_t^{-1}(p)-\Sigma _t(p)$, the (\ref{k9}) becomes an
integro-differential equation for the mass operator $\Sigma _t(p)$%
\begin{eqnarray}
\Sigma _t(p)=\frac{i\mathcal{G}}{2(2\pi )^4}\int d^4k\ \text{Tr}\left[ \frac{%
k^\alpha \gamma _\alpha +\Sigma _t(k)}{k^2-\Sigma _t^2(k)}\right] .
\end{eqnarray}
Because of momentum independence, $\Sigma _t(p)$ can now be written as $m_t$%
. This fact transforms the last equation in a self-consistent condition that
will be frequently used in the following analysis of the bound states 
\begin{eqnarray}
1=\frac{2i\mathcal{G}}{(2\pi )^4}\int \frac{d^4k}{k^2-m_t^2}.  \label{k8}
\end{eqnarray}
After regularization, (\ref{k8}) becomes an important expression linking
together cut off, dynamical mass and 4-F coupling constant $\mathcal{G}$%
\begin{eqnarray}
\mathcal{G}^{-1}=\frac{\Lambda ^2}{8\pi ^2}\left[ 1-\frac{m_t^2}{\Lambda ^2}%
\ln \left( \frac{\Lambda ^2}{m_t^2}\right) \right] .  \label{g1}
\end{eqnarray}
This is the first important result and needs some observations. First it
contains two cut off dependent terms, one logarithmically and the other
quadratically divergent. In some way it resembles the quadratic divergence
of the Higgs mechanism where a fine tuning of the Yukawa and scalar quartic
coupling constants were required. Second, for positive dangerous values of $%
m_t^2$ it implies that $\mathcal{K}=\frac{\mathcal{G}}{\mathcal{G}_c}\geq 1$
giving to $\mathcal{G}_c=\frac{8\pi ^2}{\Lambda ^2}$ the significance of a
critical coupling constant, dividing the symmetrical from the asymmetrical
phase\footnote{%
For what concerns $G_b$, its solution here is not considered for two main
reasons: the dependence on powers of $\frac 1{N_c}$ is of higher order
respect that of $G_t$ and the experimental values of $m_t$ and $m_b$ suggest
to work in the condition of maximal isospin violation.}. This means that
even the (\ref{g1}) needs a fine tuning process in order to respect the
hierarchy of scales $m_t\ll \Lambda $, together with the continuum limit $%
\stackunder{\frac{\Lambda ^2}{m_t^2}\rightarrow \infty }{\lim }\mathcal{K}=1$%
. From this point of view the spontaneous and dynamical symmetry breaking
mechanisms resolve their divergence problems in a similar way with the only
difference that from the point of view of renormalization group the
continuum limit for $\mathcal{K}$ can be interpreted as an ultraviolet
stable point 
\[
\stackunder{\frac{\Lambda ^2}{m_t^2}\rightarrow \infty }{\lim }\beta _{%
\mathcal{K}}=\frac{d\mathcal{K}}{d\ln \Lambda }=0.
\]

\section{Vertex operators}

The breaking of chiral invariance, just described, is an indispensable
ingredient for the development of the entire model. In particular, as
already stated, the use of the (\ref{k8}) in the (\ref{k5}) eliminates
quadratic divergences from vertex amplitudes and makes  possible the
individuation of poles in their spectrum. Depending on the interaction form
there are three kinds of vertices: a scalar, a pseudoscalar one and two
mixed. In all the typologies the basic instrument of research is the BS
equation. In fact the vertices operators can be written as 
\begin{eqnarray*}
\widetilde{\Gamma }_S(x,y) &=&\frac{\mathcal{G}}{2N_C}\left\langle 0\right|
T\left[ \overline{t}(0)t(0)t(x)\overline{t}(y)\right] \left| 0\right\rangle
_c=-G_{c;\alpha \beta ,\gamma \delta }^{(2)}(xy;00)\frac{\mathcal{G}}{2N_C}%
\delta _{\gamma \delta } \\
\widetilde{\Gamma }_{PS}(x,y) &=&\frac{\mathcal{G}}{2N_C}\left\langle
0\right| T\left[ \overline{t}(0)\gamma _5t(0)t(x)\overline{t}(y)\right]
\left| 0\right\rangle _c=-G_{c;\alpha \beta ,\gamma \delta }^{(2)}(xy;00)%
\frac{\mathcal{G}}{2N_C}\left( \gamma _5\right) _{\gamma \delta }
\end{eqnarray*}
\begin{eqnarray*}
\widetilde{\Gamma }_{M_{+}} &=&\frac{\mathcal{G}}{4N_C}\left\langle 0\right|
T\left[ \overline{b}(1+\gamma _5)tb\overline{t}\right] \left| 0\right\rangle
_c=-G_{c;\alpha \beta ,\gamma \delta }^{(2)}(xy;00)\frac{\mathcal{G}}{4N_C}%
\left( 1+\gamma _5\right) _{\gamma \delta } \\
\widetilde{\Gamma }_{M_{-}} &=&\frac{\mathcal{G}}{4N_C}\left\langle 0\right|
T\left[ \overline{t}(1-\gamma _5)bt\overline{b}\right] \left| 0\right\rangle
_c=-G_{c;\alpha \beta ,\gamma \delta }^{(2)}(xy;00)\frac{\mathcal{G}}{4N_C}%
\left( 1-\gamma _5\right) _{\gamma \delta }
\end{eqnarray*}
so that their Fourier transforms can be inserted in the (\ref{k5}). Again
the principal dynamical informations that enter in the equations are
collected in $\Gamma _2$ in the form of a second derivative kernel. The
problem, here, resides in the fact that the higher order in powers of $\frac
1{N_c}$ is not easily given by the first graphs of the loop expansion of $%
\Gamma _2$. In fact its double derivative carries extra $\frac 1{N_c}$
factors that forced us to take into account an infinite series of graphs
with the consequence of complicate combinatorial calculus. Moreover we
should take great care of the spinorial indices that often are the keys of
the right approximation. For reasons of convenience is preferable to treat
the scalar and pseudoscalar vertices independently from the mixed ones. Thus
we have for the kernel 
\begin{eqnarray}
\mathsf{K}_{\alpha \beta ,\gamma \delta } &=&-i\frac{\delta ^2\Gamma
_2(G_t,G_b)}{\delta G_{t;\alpha \beta }\delta G_{t;\gamma \delta }}=i\frac{%
\mathcal{G}}{2N_C}\left[ \delta _{\alpha \beta }\delta _{\gamma \delta
}+\left( i\gamma _5\right) _{\alpha \beta }\left( i\gamma _5\right) _{\gamma
\delta }\right] +  \nonumber  \label{g2} \\
&&\ \frac{i\mathcal{G}}{2N_C}\left\{ \delta _{\delta \alpha }\delta _{\beta
\gamma }\left[ 1+\sum_{n=1}^\infty \left[ F_S\right] ^n\right] +\left(
i\gamma _5\right) _{\delta \alpha }\left( i\gamma _5\right) _{\beta \gamma
}\left[ 1+\stackunder{n=1}{\sum^\infty }\left[ F_{PS}\right] ^n\right]
\right\} +  \nonumber \\
&&+0\left( \frac 1{N_C^2}\right)   \label{g2}
\end{eqnarray}
with 
\begin{eqnarray}
F_S\left( P\right)  &=&i\frac{\mathcal{G}}{2\left( 2\pi \right) ^4}\int
d^4q\ \text{Tr}\left[ G_t\left( q+\frac P2\right) G_t\left( q-\frac
P2\right) \right]   \label{g3} \\
F_{PS}\left( P\right)  &=&-i\frac{\mathcal{G}}{2\left( 2\pi \right) ^4}\int
d^4q\ \text{Tr}\left[ \gamma _5G_t\left( q+\frac P2\right) \gamma
_5G_t\left( q-\frac P2\right) \right] .  \nonumber
\end{eqnarray}
The next step, after the insertion of $\mathsf{K}_{\alpha \beta ,\gamma
\delta }$, consists in an infinite reiteration process of the BS equations.
The spinorial indices play, in this case, a major role in selecting which
reiterated terms must be kept or just ignored. The point of the question
resides in the rising of spinorial traces of propagators, each carrying an $%
N_c$ factor with it. Thus only the first term of $\mathsf{K}_{\alpha \beta
,\gamma \delta }$ is kept producing a great simplification of the final
expressions 
\begin{eqnarray}
\Gamma _{S;\alpha \beta }(P) &=&-\frac{\mathcal{G}}{2N_C}\delta _{\alpha
\beta }\left\{ 1+\stackunder{n=1}{\sum^\infty }\left[ F_S\QTR{sl}{(}P%
\QTR{sl}{)}\right] ^n\right\}  \\
\Gamma _{PS;\alpha \beta }(P) &=&-\frac{\mathcal{G}}{2N_C}\left( \gamma
_5\right) _{\alpha \beta }\left\{ 1+\stackunder{n=1}{\sum^\infty }\left[
F_{PS}\QTR{sl}{(}P\QTR{sl}{)}\right] ^n\right\} .  \nonumber
\end{eqnarray}
After regularization, it is always possible to locate a set of values of $P^2
$ for which $F\left( P\right) <1$, the last expressions becoming a
geometrical series and so can be formally rewritten as 
\begin{eqnarray}
\Gamma _{S;\alpha \beta }(P) &=&-\frac{\mathcal{G}}{2N_C}\delta _{\alpha
\beta }\left[ 1-F_S\left( P\right) \right] ^{-1}  \label{g3b} \\
\Gamma _{PS;\alpha \beta }(P) &=&-\frac{\mathcal{G}}{2N_C}\left( \gamma
_5\right) _{\alpha \beta }\left[ 1-F_{PS}(P)\right] ^{-1}.  \nonumber
\end{eqnarray}
Calculating (\ref{g3}) explicitly it is easy to express (\ref{g3b})
suitably, showing they contain a term equal to the self-consistent equation (%
\ref{k8}), and arrive to the final expressions for the vertex functions 
\begin{eqnarray}
\Gamma _S\left( P\right)  &=&-\frac i{2N_C}\left[ \frac{\left(
4m_t^2-P^2\right) }{\left( 2\pi \right) ^4}\int d^4k\ \frac 1{\left[ \left(
k+P\right) ^2-m_t^2\right] \left[ k^2-m_t^2\right] }\right] ^{-1}  \nonumber
\\
\Gamma _{PS}(P) &=&-\frac{i\gamma _5}{2N_C}\left[ \frac{P^2}{\left( 2\pi
\right) ^4}\int d^4k\ \frac 1{\left[ \left( k+P\right) ^2-m_t^2\right]
\left[ k^2-m_t^2\right] }\right] ^{-1}.  \label{g4}
\end{eqnarray}
A similar procedure takes to the evaluation of the mixed vertex functions.
The differences, in this case, are a slight change of spinor indices in the
BS equation and a more complicate kernel 
\[
\mathsf{K}_{\alpha \beta ,\gamma \delta }\left[ q;\left( p-k\right) \right]
=-\frac{i\mathcal{G}}{4N_C}\left( 1+\gamma _5\right) _{\delta \alpha }\left(
1-\gamma _5\right) _{\beta \gamma }\left\{ 1+\stackunder{n=1}{\sum^\infty }%
\left[ F_M\left( P\right) \right] ^n\right\} 
\]
leading to 
\begin{eqnarray}
\Gamma _{M_{\pm };\alpha \beta }(P) &=&-\frac{\mathcal{G}}{4N_C}\left( 1\pm
\gamma _5\right) _{\alpha \beta }-i\frac{\mathcal{G}}{4N_C}\frac{\left( 1\pm
\gamma _5\right) _{\alpha \beta }}{(2\pi )^4}\left\{ 1+\stackunder{n=1}{%
\sum^\infty }\left[ F_M\left( P\right) \right] ^n\right\} \times   \nonumber
\label{g4a} \\
&&\times \int d^4q\ N_C\text{Tr}\left[ \left( 1\mp \gamma _5\right)
G_t\left( q+P\right) \Gamma _{M_{\pm }}\left( P\right) G_b\left( q\right)
\right]   \label{g4a}
\end{eqnarray}
with 
\[
F_M\left( P\right) =\frac{i\mathcal{G}}{4\left( 2\pi \right) ^4}\int d^4l\ 
\text{Tr}\left[ \left( 1-\gamma _5\right) G_t\left( l+P\right) \left(
1+\gamma _5\right) G_b\left( l\right) \right] .
\]
Unfortunately the reiteration of (\ref{g4a}) doesn't eliminate any term and
produces a final expression that, only under the right decomposition,
becomes 
\begin{eqnarray}
\Gamma _{M_{\pm };\alpha \beta }(P)=-\frac{\mathcal{G}}{4N_C}\left( 1\pm
\gamma _5\right) _{\alpha \beta }\left[ \frac{F_M\left( P\right) ^2}{%
1-F_M\left( P\right) }\right]   \label{g5}
\end{eqnarray}
with 
\begin{eqnarray}
1-F_M\left( P\right)  &=&\frac 12-\frac{i\mathcal{G}}{(2\pi )^4}\int \frac{%
d^4k\ \left( k^2-P^2\right) }{\left( P+k\right) ^2\left( k^2-m_t^2\right) }%
\stackunder{P^2\rightarrow 0}{\longrightarrow }0  \label{g5a} \\
F_M\left( P\right) ^2 &=&\left\{ -\frac 12-\frac{i\mathcal{G}}{(2\pi )^4}%
\int \frac{d^4k\ \left( k^2-P^2\right) }{\left( P+k\right) ^2\left(
k^2-m_t^2\right) }\right\} ^2\stackunder{P^2\rightarrow 0}{\longrightarrow }%
1.  \nonumber
\end{eqnarray}
Collecting results from (\ref{g4}), (\ref{g5}) and (\ref{g5a}) it is
straight now to observe that the scalar vertex presents a pole for a
transferred momentum $P=2m_t$, while the pseudoscalar and mixed ones have
poles at $P^2=0$ corresponding to bosonic bound states on their mass shell.
In particular the scalar channel represents a top-antitop state that could
be compared to a sort of pseudo-Higgs boson because of its extremely massive
characteristic. Moreover the absence of a negative mass squared bound state
is a first indication of the stability of the asymmetrical phase. The
natural conclusion is that the pseudoscalar and mixed resonances are a sort
of pseudo-Goldstone bosons coming from the breaking of the chiral
invariance. At this point the logic hope is that these latter amplitudes
shall be absorbed in some ways in the gauge electroweak propagators in the
form of mass terms. The analogy with the Higgs mechanism would be, then,
complete.

\section{Electroweak symmetry breaking}

Having all the basic prerequisites in our hand it is now possible to
approach the main target of this work. Because the SD equation coming from
the stationary condition of the effective action is the main tool used in
here, let us recall the loop expansion in terms of gauge propagators 
\begin{eqnarray}
\Gamma (\mathcal{D}_{\mu \nu },G_t,G_b) &=&\frac i2\sum_{i=W,A,Z^0}\left[
\iint d^4x\ d^4y\ \text{Tr}\left\{ \mathrm{\ln }\left[ D_{\mu \nu }^i\left(
x;y\right) \mathcal{D}_{\mu \nu }^{i\ ^{-1}}\left( y;x\right) \right]
+\right. \right.   \nonumber \\
&&\ \ \ \ \ +\left. \left. D_{\mu \nu }^{i\ ^{-1}}\left( x;y\right) \mathcal{%
D}_{\mu \nu }^i\left( y;x\right) -1\right\} +\Gamma _2(\mathcal{D}_{\mu \nu
}^i,G_t,G_b)\right] .  \label{g6}
\end{eqnarray}
We obviously didn't mention terms dependent exclusively on fermion
propagators because they would be eliminated by the stationary condition
after the derivation respect $\mathcal{D}_{\mu \nu }^i$. Again the richness
of the theory is enclosed in $\Gamma _2$ and so depends on the form of
interaction terms. In this case such terms are not only given by the 4-F
interaction, but depends directly on the form of the neutral and charged
currents $\mathbf{J}_3^\mu =\frac 14\left[ \overline{t}\gamma ^\mu \left(
1-\gamma _5\right) t-\overline{b}\gamma ^\mu \left( 1-\gamma _5\right)
b\right] $, $\mathbf{J}_Y^\mu =\left[ \frac 23\left( \overline{t}\gamma ^\mu
t\right) -\frac 13\left( \overline{b}\gamma ^\mu b\right) -J_3^\mu \right] $
and $\mathbf{J}^\mu =\left[ \overline{b}\left( 1+\gamma _5\right) \gamma
^\mu t\right] $ with its complex conjugate. Consequently the combinatorics
of 2PI vacuum graphs becomes more complicate and necessitates an accurate
evaluation that keeps an infinite number of terms of the same order in power
of $N_c$. Taking into account only the higher order terms, the final
expression, in the case of charged weak propagators is 
\begin{eqnarray}
\Gamma _2\left( \mathcal{D}_{\mu \nu }^W,G_t,G_b\right)  &=&-\frac{g_2^2}{16}%
N_C\iint d^4x\ d^4y\ \left\{ \text{Tr}\left[ G_t\left( y;x\right) \gamma
^\nu \left( 1-\gamma _5\right) \right. \right. \times   \nonumber  \label{g7}
\\
&&\ \ \times \left. \left. G_b\left( x;y\right) \left( 1+\gamma _5\right)
\gamma ^\mu \right] \mathcal{D}_{\mu \nu }^W\left( x;y\right) \right\} + 
\nonumber  \label{g7} \\
&&\ \ -\frac{ig_2^2\mathcal{G}}{64}N_C\left[ \stackunder{n=1}{\sum^\infty }%
F_W^{\left( n\right) }\left( \mathcal{D}_{\mu \nu }^W,G_t,G_b\right) \right] 
\label{g7}
\end{eqnarray}
with 
\begin{eqnarray*}
F_W^{\left( n\right) } &=&\left( \frac{i\mathcal{G}}4\right) ^{n-1}\idotsint
d^4xd^4z_1\ldots d^4z_nd^4y\left\{ \text{Tr}\left[ G_b\left( x;z_1\right)
\left( 1+\gamma _5\right) G_t\left( z_1;x\right) \right. \right. \times  \\
&&\ \ \times \left. \gamma ^\nu \left( 1-\gamma _5\right) \right] \times 
\text{Tr}\left[ G_b\left( z_2;z_1\right) \left( 1-\gamma _5\right) G_t\left(
z_1;z_2\right) \left( 1+\gamma _5\right) \right] \times \ldots  \\
&&\ \ \ldots \times \left. \text{Tr}\left[ G_b\left( z_n;z_{n-1}\right)
\left( 1-\gamma _5\right) G_t\left( z_{n-1};z_n\right) \left( 1+\gamma
_5\right) \right] \right. \times  \\
&&\ \ \left. \text{Tr}\left[ G_t\left( y;z_n\right) \left( 1+\gamma
_5\right) G_b\left( z_n;y\right) \gamma ^\mu \left( 1-\gamma _5\right)
\right] \mathcal{D}_{\mu \nu }^W\left( x;y\right) \right\} .
\end{eqnarray*}
Applying the stationary condition $\frac{\delta \Gamma \left( \mathcal{D}%
_{\mu \nu }^W,G_t,G_b\right) }{\delta \mathcal{D}_{\mu \nu }^W}=0$ to the (%
\ref{g6}) one directly obtains the SD equation for the complete propagator $%
\mathcal{D}_{\mu \nu }^W$ as a complicate function of the free one $D_{\mu
\nu }^W$ and the fermionic $G_t$, $G_b$, that demonstrate the complete
absorption of the mixed vertex functions 
\begin{eqnarray}
\mathcal{D}_{\mu \nu }^{W\ ^{-1}} &=&D_{\mu \nu }^{W^{\ -1}}+\frac{iN_Cg_2^2}%
8\left\{ \text{Tr}\left[ G_t\gamma ^\nu \left( 1-\gamma _5\right) G_b\left(
1+\gamma _5\right) \gamma ^\mu \right] \right. +iR\left( F_M\left( P\right)
\right) \times   \nonumber \\
&&\ \ \ \ \ \times \left. \text{Tr}\left[ G_t\gamma ^\nu \left( 1-\gamma
_5\right) G_b\Gamma _{M_{+}}\right] \text{Tr}\left[ \Gamma
_{M_{-}}G_tG_b\gamma ^\mu \left( 1-\gamma _5\right) \right] \right\} .
\label{g7b}
\end{eqnarray}
After a long and accurate regularization \cite{b2} that doesn't break the
gauge invariance one recover an expression that makes evident how the
absorption process is a concrete ''mass generation'' mechanism giving body
to the electroweak symmetry breaking 
\begin{eqnarray}
\frac 1{g_2^2}\mathcal{D}_{\mu \nu }^{W\ ^{-1}}=i\left( \frac{P^\mu P^\nu }{%
P^2}-g^{\mu \nu }\right) \left( \frac 1{\widetilde{g}_2^2\left( P^2\right)
}P^2-\widetilde{h}^2\left( P^2\right) \right)   \label{g8}
\end{eqnarray}
with 
\begin{eqnarray*}
\frac 1{\widetilde{g}_2^2\left( P^2\right) } &=&\frac 1{g_2^2}+\frac{N_C}{%
48\pi ^2}\left[ \ln \frac{\Lambda ^2}{m_t^2}+P^2\int_{m_t^2}^\infty \frac{%
d^4k}{k^2\left( k^2-P^2\right) }\left( 1-\frac{m_t^2}{k^2}\right) ^2\left( 1+%
\frac{2m_t^2}{k^2}\right) \right]  \\
\widetilde{h}^2\left( P^2\right)  &=&\frac{N_Cm_t^2}{32\pi ^2}\left[ \ln 
\frac{\Lambda ^2}{m_t^2}+P^2\int_{m_t^2}^\infty \frac{d^4k}{k^2\left(
k^2-P^2\right) }\left( 1-\frac{m_t^2}{k^2}\right) \right] .
\end{eqnarray*}
The (\ref{g8}) has, evidently, a pole in correspondence of $\mathbf{M}%
_W^2=P^2=\widetilde{g}_2^2\left( P^2\right) \widetilde{h}^2\left( P^2\right) 
$ that is just the gauge mass term searched.

A similar procedure, even if a bit more lengthy, applies for the neutral
gauge propagators. After the additional transform that rotate fields from $%
\mathbf{W}_{3\mu },\mathbf{B}_\mu $ to $\mathbf{Z}_\mu ^0,\mathbf{A}_\mu $
changing the interaction terms, the relevant difference is contained in the
explicit expressions of the $\Gamma _2$ expansion 
\begin{eqnarray}
\Gamma _2=\Gamma _2^A+\Gamma _2^{Z^0}
\end{eqnarray}
where the photonic part is 
\begin{eqnarray}
\Gamma _2^A &=&-\frac 29e^2N_C\iint d^4x\ d^4y\ \left\{ \text{Tr}\left[
G_t\left( y;x\right) \gamma ^\mu G_t\left( x;y\right) \gamma ^\nu \right] 
\mathcal{D}_{\mu \nu }^A\left( x;y\right) \right\} +  \nonumber \\
&&\ -\frac{ie^2\mathcal{G}}9N_C\left\{ \left[ \stackunder{n=1}{\sum^\infty }%
F_{A,S}^{\left( n\right) }\right] +\left[ \stackunder{n=1}{\sum^\infty }%
F_{A,PS}^{\left( n\right) }\right] \right\}
\end{eqnarray}
with 
\begin{eqnarray*}
F_{A,S}^{\left( n\right) } &=&\left( \frac{i\mathcal{G}}2\right)
^{n-1}\idotsint d^4xd^4z_1\ldots d^4z_nd^4y\left\{ \text{Tr}\left[ G_t\left(
x;z_1\right) \gamma ^\mu G_t\left( z_1;x\right) \right] \right. \times \\
&&\ \times \text{Tr}\left[ G_t\left( z_2;z_1\right) G_t\left( z_1;z_2\right)
\right] \ldots \text{Tr}\left[ G_t\left( z_n;z_{n-1}\right) G_t\left(
z_{n-1};z_n\right) \right] \times \\
&&\ \times \left. \text{Tr}\left[ G_t\left( y;z_n\right) G_t\left(
z_n;y\right) \gamma ^\nu \right] \mathcal{D}_{\mu \nu }^A\left( x;y\right)
\right\}
\end{eqnarray*}
\begin{eqnarray*}
F_{A,PS}^{\left( n\right) } &=&\left( \frac{i\mathcal{G}}2\right)
^{n-1}\idotsint d^4xd^4z_1\ldots d^4z_nd^4y\left\{ \text{Tr}\left[ G_t\left(
x;z_1\right) \gamma ^\mu G_t\left( z_1;x\right) \gamma _5\right] \right.
\times \\
&&\ \times \text{Tr}\left[ \gamma _5G_t\left( z_2;z_1\right) \gamma
_5G_t\left( z_1;z_2\right) \right] \ldots \text{Tr}\left[ \gamma _5G_t\left(
z_n;z_{n-1}\right) \gamma _5G_t\left( z_{n-1};z_n\right) \right] \times \\
&&\ \times \left. \text{Tr}\left[ G_t\left( y;z_n\right) \gamma _5G_t\left(
z_n;y\right) \gamma ^\nu \right] \mathcal{D}_{\mu \nu }^A\left( x;y\right)
\right\}
\end{eqnarray*}
meanwhile the $Z^0$ part has the following form 
\begin{eqnarray}
\Gamma _2^{Z^0} &=&-\frac{g_2^2N_C}{32\cos ^2\theta _W}\iint d^4xd^4y\
\left\{ \text{Tr}\left[ G_t\left( y;x\right) \gamma ^\mu \left( \left(
1-\frac 83\sin ^2\theta _W\right) -\gamma _5\right) \right. \right. \times 
\nonumber \\
&&\ \times \left. \left. G_t\left( x;y\right) \left( \left( 1-\frac 83\sin
^2\theta _W\right) -\gamma _5\right) \right] \mathcal{D}_{\mu \nu
}^{Z^0}\right\} -\frac{ig_2^2\mathcal{G}}{64\cos ^2\theta _W}N_C\times 
\nonumber \\
&&\ \times \left\{ \left[ \stackunder{n=1}{\sum^\infty }F_{Z^0,S}^{\left(
n\right) }\right] +\left[ \stackunder{n=1}{\sum^\infty }F_{Z^0,PS}^{\left(
n\right) }\right] \right\}
\end{eqnarray}
with 
\begin{eqnarray*}
F_{Z^0,S}^{\left( n\right) } &=&\left( \frac{i\mathcal{G}}2\right)
^{n-1}\idotsint d^4xd^4z_1\ldots d^4z_nd^4y\ \mathcal{D}_{\mu \nu
}^{Z^0}\left( x;y\right) \times \\
&&\times \left\{ \text{Tr}\left[ G_t\left( x;z_1\right) \gamma ^\mu \left(
\left( 1-\frac 83\sin ^2\theta _W\right) -\gamma _5\right) G_t\left(
z_1;x\right) \right] \right. \\
&&\times \text{Tr}\left[ G_t\left( z_2;z_1\right) G_t\left( z_1;z_2\right)
\right] \ldots \text{Tr}\left[ G_t\left( z_n;z_{n-1}\right) G_t\left(
z_{n-1};z_n\right) \right] \times \\
&&\times \left. \text{Tr}\left[ G_t\left( y;z_n\right) G_t\left(
z_n;y\right) \gamma ^\nu \left( \left( 1-\frac 83\sin ^2\theta _W\right)
-\gamma _5\right) \right] \right\}
\end{eqnarray*}
\begin{eqnarray*}
F_{Z^0,PS}^{\left( n\right) } &=&\left( \frac{i\mathcal{G}}2\right)
^{n-1}\idotsint d^4xd^4z_1\ldots d^4z_nd^4y\ \mathcal{D}_{\mu \nu
}^{Z^0}\left( x;y\right) \times \\
&&\times \left\{ \text{Tr}\left[ G_t\left( x;z_1\right) \gamma ^\mu \left(
\left( 1-\frac 83\sin ^2\theta _W\right) -\gamma _5\right) G_t\left(
z_1;x\right) \gamma _5\right] \right. \\
&&\times \text{Tr}\left[ \gamma _5G_t\left( z_2;z_1\right) \gamma
_5G_t\left( z_1;z_2\right) \right] \ldots \text{Tr}\left[ \gamma _5G_t\left(
z_n;z_{n-1}\right) \gamma _5G_t\left( z_{n-1};z_n\right) \right] \times \\
&&\times \left. \text{Tr}\left[ G_t\left( y;z_n\right) \gamma _5G_t\left(
z_n;y\right) \gamma ^\nu \left( \left( 1-\frac 83\sin ^2\theta _W\right)
-\gamma _5\right) \right] \right\} .
\end{eqnarray*}
Again considering the stationary conditions for the neutral gauge boson
propagators lead us to 
\begin{eqnarray}
\mathcal{D}_{\mu \nu }^{A^{\ -1}}\left( P\right) &=&D_{\mu \nu }^{A^{\
-1}}\left( P\right) +\frac{4iN_Ce^2}9\left\{ \text{Tr}\left[ G_t\gamma ^\mu
G_t\gamma ^\nu \right] \right. +  \nonumber \\
&&-i\Gamma _S\text{Tr}\left[ G_t\gamma ^\mu G_t\right] \text{Tr}\left[
G_t\gamma ^vG_t\right] +  \nonumber \\
&&-\left. i\Gamma _{PS}\text{Tr}\left[ G_t\gamma _5G_t\gamma ^\mu \right] 
\text{Tr}\left[ G_t\gamma _5G_t\gamma ^\nu \right] \right\}  \label{g9}
\end{eqnarray}
and 
\begin{eqnarray}
\mathcal{D}_{\mu \nu }^{Z^{0\ -1}}\left( P\right) &=&D_{\mu \nu }^{Z^{0\
-1}}\left( P\right) +\frac{iN_Cg_2^2}{16\cos ^2\theta _W}\left\{ \text{Tr}%
\left[ G_t\gamma ^\mu \left( \left( 1-\frac 83\sin ^2\theta _W\right)
-\gamma _5\right) \right. \right. \times  \nonumber \\
&&\times \left. G_t\gamma ^\nu \left( \left( 1-\frac 83\sin ^2\theta
_W\right) -\gamma _5\right) \right] +  \nonumber \\
&&-i\Gamma _S\text{Tr}\left[ G_tG_t\gamma ^\mu \left( \left( 1-\frac 83\sin
^2\theta _W\right) -\gamma _5\right) \right] \times  \nonumber \\
&&\times \text{Tr}\left[ G_tG_t\gamma ^v\left( \left( 1-\frac 83\sin
^2\theta _W\right) -\gamma _5\right) \right] +  \nonumber \\
&&-i\Gamma _{PS}\text{Tr}\left[ G_t\gamma _5G_t\gamma ^\mu \left( \left(
1-\frac 83\sin ^2\theta _W\right) -\gamma _5\right) \right] \times  \nonumber
\\
&&\times \left. \text{Tr}\left[ G_t\gamma _5G_t\gamma ^{Gv}\left( \left(
1-\frac 83\sin ^2\theta _W\right) -\gamma _5\right) \right] \right\}
\label{h1}
\end{eqnarray}
giving expressions in which vertex functions appear naturally. Explicit
calculations bring us to 
\begin{eqnarray}
\frac 1{g_2^2}\mathcal{D}_{\mu \nu }^{Z^{0\ -1}}\left( P\right) =i\left( 
\frac{P^\mu P^\nu }{P^2}-g^{\mu \nu }\right) \left( \frac 1{\overline{g}%
_2^2\left( P^2\right) }P^2-\widetilde{f}^2\left( P^2\right) \right)
\end{eqnarray}
with 
\begin{eqnarray*}
\frac 1{\overline{g}_2^2\left( P^2\right) } &=&\frac 1{g_2^2}+\frac{N_C}{%
96\pi ^2\cos ^2\theta _W}\left( \left( 1-\frac 83\sin ^2\theta _W\right)
^2+1\right) \left[ \ln \frac{\Lambda ^2}{m_t^2}\right. + \\
&&+\left. P^2\int_{4m_t^2}^\infty \frac{d^4k}{k^2\left( k^2-P^2\right) }%
\left( 1-\frac{4m_t^2}{k^2}\right) ^{\frac 12}\left( 1+\frac{2m_t^2}{k^2}%
\right) \right] \\
\widetilde{f}^2\left( P^2\right) &=&\frac{N_Cm_t^2}{32\pi ^2\cos ^2\theta _W}%
\left[ \ln \frac{\Lambda ^2}{m_t^2}+\frac{P^2}2\int_{4m_t^2}^\infty \frac{%
d^4k}{k^2\left( k^2-P^2\right) }\left( 1-\frac{4m_t^2}{k^2}\right) ^{-\frac
12}\right]
\end{eqnarray*}
giving to the $\mathbf{Z}^0$ field a mass term $\mathbf{M}_{Z^0}^2=P^2=%
\widetilde{g}_2^2\left( P^2\right) \widetilde{f}^2\left( P^2\right) $.
Meanwhile for the photon propagator one obtains naturally no mass correction
but only a coupling constant renormalization. 
\begin{eqnarray}
\frac 1{e^2}\mathcal{D}_{\mu \nu }^{A^{\ -1}}\left( P\right) =i\left( P^\mu
P^\nu -g^{\mu \nu }P^2\right) \left( \frac 1{\widetilde{e}^2\left(
P^2\right) }\right)
\end{eqnarray}
with 
\[
\frac 1{\widetilde{e}^2\left( P^2\right) }=\frac 1{e^2}+\frac{2N_C}{27\pi ^2}%
\left[ \ln \frac{\Lambda ^2}{m_t^2}+P^2\int_{4m_t^2}^\infty \frac{d^4k}{%
k^2\left( k^2-P^2\right) }\left( 1-\frac{4m_t^2}{k^2}\right) ^{\frac
12}\left( 1+\frac{2m_t^2}{k^2}\right) \right] . 
\]
As we have already noted, in (\ref{g7b}), (\ref{g9}) and (\ref{h1}) the
inverse gauge propagators are corrected by means of terms that involve
vertex functions. In a Feynman graph representation these terms contain
internal lines corresponding to the resonance channels found in the section
4. In other words corrections to the gauge propagators imply inevitably the
exchange of bound state particles. Moreover calculation details show \cite
{b2} that the only vertices that contribute to the mass generation are those
relative to the pseudo-Goldstone modes, while the scalar amplitude remains
outside the process: a strong hint towards the identification of latter's
resonance with a composite Higgs boson.

\section{General considerations and conclusions}

This work certainly doesn't constitute the only attempt to build a model
that try to justify the largeness of the quark top mass and use it as a
breakthrough in the electroweak symmetry breaking. An entire class of so
called ''Top Mode Standard Model'' (TMSM) were written in the years that
divide the last two decades\cite{b4}. A second phase for TMSM arrived after
a period of five years when a series of works tried to improve the mechanism
giving to it the label ''renormalizable''\cite{b6}\cite{b7}\cite{b75}.
Despite their results and conclusions on the gauged NJL model and its
renormalization, I would like to emphasize some aspects of the approach
exposed here. Forgetting for the moment the divergencies included in the
self-consistent equation (\ref{g1}), I show the importance of the chiral
symmetry breaking as a means through which obtain a gauge symmetry breaking.
The appearance of corrections depending on bound states amplitudes, in the
SD equations for the inverse gauge propagators, makes the dynamical
''absorption process'' a real alternative model to the static Higgs
mechanism. This result is not conclusive obviously: the expressions obtained
in section 5 gives in terms of $\frac 1{N_c}$ power series the following
behavior 
\[
\mathcal{D}_{\mu \nu }=0\left( 1\right) +\left( \frac 1{N_c}\right) +\ldots
. 
\]
This implies that in the SD equation for $G_t$ the neglect of boson
contributions to $\Gamma _2$ wasn't the correct procedure. At least we
should take in consideration the gauge propagator contribution to graphs in
planar approximation which, in fact, carries the same order (in $\frac
1{N_c} $) of the 4-F interaction. This was not done for two main reasons:
the preponderance of 4-F induced terms due to the overcritical value of $%
\mathcal{G}$ respect to $g_1$ and $g_2$. Second, because the 4-F hypothesis
was analyzed in a low-energy regime (the SD without gauge contributions
could be considered the zero approximation for $m_t$ in a $P^2$ power series
expansion) the ''generation'' of gauge masses would be then justified as a
tree-level approximation. In this sense the next logical step would be the
inclusion of gauge corrections to all the equations of the model. The
program is compatible with fermionic SD and BS equations, but becomes really
complex for gauge SD equations because of the combinatorics of the graphs
which clearly complicate the localization of the ''generated'' mass terms.

Second important remark concerns the supposed predictability this model can
give of the SM mass spectrum. In section 3, in fact, I ignored some
interaction terms in order to simplify the analysis of the successive
sections. Even if not mentioned here I have studied \cite{b2} the
consequences, on the fermion mass sector, of the insertion of additional 4-F
interactions in the Lagrangian. In analogy with the work of Hasenfratz et
Al. \cite{b11}, I found that on this side gauged NJL models don't give any
additional previsional information compared with the traditional Higgs
mechanism. This should not divert the attention from the fundamental
dynamical content that could help to enlighten several aspects of existing
problems. The NJL model is still a fruitful argument of analysis and will be
again a starring of future researches.

\end{document}